\documentstyle[12pt]{article}

\thicklines                         
\def\a{\alpha}
\def\b{\beta}

\def\d{\delta}
\def\e{\epsilon}                
\def\g{\gamma}

\def\j{\psi}

\def\m{\mu}

\def\o{\omega}
\def\p{\pi}                     
\def\th{\theta}                  
\def\r{\rho}                    
\def\s{\sigma}                  

\def\x{\xi}

\def\D{\Delta}
\def\F{\Phi}

\def\O{\Omega}

\def\cb{{\cal B}}
\def\cc{{\cal C}}
\def\cd{{\cal D}}

\def\ch{{\cal H}}   

\def\co{{\cal O}}
\def\cp{{\cal P}}

\def\car{{\cal R}}

\def\ct{{\cal T}}

\def\cbo{{\,\raise-.15ex\Sc [\,}}                       

\def\bra#1{\Big\langle #1\Big|}                 
\def\ket#1{\Big| #1\Big\rangle}                 
\def\sbra#1{\left\langle #1\right|}             
\def\sket#1{\left| #1\right\rangle}             
\def\svev#1{\left\langle #1\right\rangle}       
\def\ddt#1{{\buildrel {\hbox{\LARGE .\kern-2pt.}} \over {#1}}}

\def\beqn#1{ \renewcommand{\theequation}{#1}
             \begin{equation} }
\def\eeqn{\end{equation}}
\def\beqr#1{ \renewcommand{\theequation}{#1}
             \begin{eqnarray} }
\def\eeqr{\end{eqnarray}}
\def\NON{\nonumber\\}
\def\beqrabc#1{ \setcounter{equation}{0}
                \renewcommand{\theequation}{#1\alph{equation}}
                \begin{eqnarray} }
\def\beqrn#1#2{ \setcounter{equation}{#2}
                \renewcommand{\theequation}{#1.\arabic{equation}}
                \begin{eqnarray} }

\def\APH#1{Ann. Phys. {\bf #1}}

\def\NPB#1{Nucl. Phys. {\bf B#1}}
\def\NPBP#1{Nucl. Phys. (Proc. Suppl.) {\bf B#1}}
\def\PLB#1{Phys. Lett. {\bf B#1}}
\def\PRD#1{Phys. Rev. {\bf D#1}}
\def\PR#1{Phys. Rev. {\bf #1}}
\def\PRL#1{Phys. Rev. Lett. {\bf #1}}

\def\APP#1{Acta. Phys. Pol. {\bf B#1}}

\def\MPL#1{Mod. Phys. Lett. {\bf A #1}}

\def\sstyle{\scriptstyle}

\def\rhs{\mbox{r.h.s.} }

\def\ie{\mbox{i.e.} }
\def\eg{\mbox{e.g.} }

\def\Int#1{\int\! d#1\,}
\def\dInt#1#2#3{\int_{#1}^{#2}d#3\,}
\def\frac#1#2{ {\sstyle {#1\over #2} } }
\def\det#1{{\rm det}\left(#1\right)}

\def\half{{1\over 2}}
\def\Re{{\rm Re\,}}
\def\Im{{\rm Im\,}}

\def\det{{\rm det\,}}

\def\sgmdt#1{\,\mbox{\boldmath $\s\cdot$} {\bf #1}}

\begin{document}
\hfill WIS--93/57--JULY--PH
\par
\begin{center}
\vspace{15mm}
{\large\bf On the Absence of Chiral Fermions\\
in Interacting Lattice Theories}\\[5mm]
{\it by}\\[5mm]
Yigal Shamir\\
Department of Physics\\
Weizmann Institute of Science, Rehovot 76100, ISRAEL\\
email: ftshamir@weizmann.weizmann.ac.il\\[15mm]
{ABSTRACT}\\[2mm]
  \end{center}
\begin{quotation}
  We consider interacting theories with a compact internal symmetry
group on a regular lattice. We show that the spectrum
is necessarily vector-like provided the
following conditions are satisfied: (a)~weak form of locality,
(b)~relativistic continuum limit without massless bosons, and
(c)~pole-free effective vertex functions for conserved currents.

  The proof exploits the zero frequency inverse retarded propagator of
an appropriate set of interpolating fields as an effective quadratic
hamiltonian, to which the Nielsen-Ninomiya theorem is applied.

\end{quotation}

\newpage
\noindent {\bf 1.~~Introduction and Conclusions}
\vspace{3ex}

  The only rigorous way, presently known to us, to define non-abelian
gauge theories, relies on the lattice as a regulator. The observed
fermion spectrum fits into a chiral representation of
SU(3)$\times$SU(2)$\times$U(1), and so the construction of a consistent
chiral gauge theory on the lattice has been a major goal in theoretical
physics.

  In spite of extensive efforts, this program
has been unsuccessful to date. The basic stumbling block is
the doubling problem~[1,2]. A naive discretization
of the continuum hamiltonian of a Weyl fermion gives rise to eight
Weyl fermions in the classical continuum limit of the lattice
hamiltonian. If one starts with a Dirac fermion, the doublers can be
eliminated by introducing the Wilson term. But the price is that the
axial symmetry of the classical continuum hamiltonian is lost.

  There is an intimate relation between species doubling on the lattice
and chiral anomalies in the continuum. Although classically conserved
for a massless Dirac fermion, the conservation of the axial current is
violated by quantum effects in the continuum theory.
On the other hand, the lattice is a physical regulator, and so
any current which is conserved at tree level
should be conserved in the continuum limit as well.
If one insists on keeping
the axial symmetry of the lattice action, the conflict between the
predictions of the lattice theory and the continuum theory is resolved
by the appearance of the doublers. The obvious price is that now, the
continuum limit of the lattice theory is different from the continuum
theory from which we have started.

  Following the work of Karsten and Smit~[3], the precise
conditions for fermion doubling in a free
fermionic theory defined on a regular lattice were stated by Nielsen and
Ninomiya  as a no-go theorem~[4]. They assume the existence of a set of
exactly conserved, locally defined charges which admit discrete
eigenvalues. The Nielsen-Ninomiya (NN) theorem then asserts that
there must be an equal
number of positive helicity and negative helicity fermions in every
complex representation of the symmetry group, provided the Fourier
transform of the free hamiltonian
has a continuous first derivative. (Recall
that chirality equal helicity for massless fermions). The
NN theorem applies in particular when the hamiltonian has
a short range, and the charges are constructed canonically and generate
a compact Lie group.

  The absence of chiral fermions is essentially a counting theorem
about the zeros of the free hamiltonian in the Brillouin zone.
It takes its simplest form
in one space dimension. A right (left) mover is an eigenvalue of
$H(p)$ which satisfies
$$
E(p)=\pm(p-p_c)+O((p-p_c)^2)\,,
\eqno(1)
$$
in the neighbourhood of some $p_c$. The  presence of an equal number
of plus and minus signs is then nothing but the familiar property that
the graph of a function from the circle into the real numbers which
has a continuous derivative and regular zeros must cross the axis
upwards and downwards an equal number of times.

   In three space dimensions, a massless fermion
corresponds to level crossing which is described by the
effective two-by-two hamiltonian
$$
H_{eff}({\bf p})=\pm\sgmdt{(p-p}_c) +
O(({\bf p - p}_c)^2)\,.
\eqno(2)
$$
The point ${\bf p}_c$ is called a {\it degeneracy point}. The $\pm$
signs correspond to the helicity of the fermion. The physically more
transparent proof of the NN theorem (second paper of ref.~[4])
relies on the examination of
trajectories of constant $\s_3$ in the energy surface, which go
through at least one degeneracy point.
Since the Brillouin zone is topologically a three-torus, these curves
must be closed. When the curve goes through a degeneracy point, it
crosses from positive to negative $E$ (or vice versa). Moreover, one
can choose an orientation to the curve using the local properties of
the wave function, and show that near a degeneracy point the
orientation (thought of as a unit tangent vector) coincides with the
helicity of the fermion. The fact that every differentiable curve
is orientable than implies that every crossing which corresponds to a
left handed particle must be followed by a crossing
through another degeneracy point which describes a right
handed particle.

  The continuum limit of asymptotically free gauge theories is
achieved at vanishing bare coupling. Together with the success of
perturbative QCD in deep inelastic scattering, this leads to the
generally accepted view that the fermionic spectrum can be correctly
determined by setting the gauge couplings to zero. In the absence of
other interactions, the NN theorem implies that the
fermionic spectrum must be vector-like provided the hamiltonian has
a short range.

  Attempts to avoid fermion doubling by using long range lattice
derivatives lead to various inconsistencies at the level of weak
coupling perturbation theory. Important examples include the
SLAC derivative~[5]  which avoids the
extra zeros by creating a discontinuity in the dispersion relation, and
a method due to Rebbi~[6] which is characterized by the presence
of a pole in the dispersion relation. The former suffers from
Lorentz violations and non-locality~[7] while the latter suffers from
the presence of ghosts~[8].

  As it stands, the NN theorem does not apply if the
lattice model contains some strong interactions. This observation have
led to several proposals~[9] for constructing chiral gauge theories
on the lattice which exploit a common strategy. (See ref.~[10]
for a review). One starts with a model
containing only fermions and (possibly) scalar fields. In addition to
standard quadratic terms, one introduces judiciously chosen strong
interactions among these fields which are operative at the lattice
scale, and vanish in the classical continuum limit.
Local symmetries of the desired continuum theory -- the target theory --
should appear at this stage as exact global symmetries of the model.

  The strong interactions introduced
at the lattice scale do not survive in the continuum limit.
The continuum limit of the fermion-scalar
model is a {\it free} theory of massless  fermions. The hope
is, that if the phase diagram is sufficiently non-trivial, arguments
based on weak coupling perturbation theory will not apply, and
a point in the phase diagram will be found in which
the all the doublers have been decoupled and the massless spectrum is
chiral. It is crucial that, at the same time, the to-be-gauged
global symmetries are not broken spontaneously.

  If this program were successful, a consistent chiral
gauge theory could be obtained by turning on the gauge interactions in
such a model. However, explicit model calculations have lead
to negative conclusions in all cases studied so far~[10,11].

  Our purpose in the present paper is to provide a general treatment of
the problem, which applies for example also to similar models
based on the recently proposed domain wall fermions~[12].
We will consider a hamiltonian defined on a regular lattice
that has a compact global symmetry group
which is not spontaneously broken.
We will prove that under mild assumptions, which are directly related
to the physical properties of a consistent continuum limit,
the spectrum is necessarily vector-like. A summary of our main results
is contained in ref.~[13]

  Our proof is based on the observation that, while in general
the physics of an interacting theory may be very complicated,
all the crucial ingredients of the NN theorem have natural
generalizations to this case. The object which plays the role of
an effective hamiltonian is the {\it zero frequency inverse propagator}.
Analyticity, or more generally, differentiability properties
of the propagator are related to the range of the hamiltonian
and hence to the {\it locality} of the theory. This relation can
be exhibited most directly if we choose to work with the {\it retarded
propagator}. Moreover, the desired properties of the continuum limit
imply that massless fermions can still be identified with the zeros
of the effective hamiltonian, and that these zeros still take the form
of eq.~(2). Finally, needless to say, periodicity across the Brillouin
zone is mandatory in the interacting theory as well.

  We first present a heuristic argument that shows how, using the
above ingredients, one can conclude that the spectrum must be
vector-like in interacting theories too. For methodological reasons,
we will base this argument on the more familiar self energy,
\ie the inverse of the time ordered propagator. We comment that,
anyhow, at zero frequency all propagators coincide because the pole
prescription is irrelevant.

  For definiteness, consider a hamiltonian defined on a cubic lattice
whose spectrum contains in particular a left handed fermion
in the vicinity of the origin of the Brillouin zone.
Suppose that this chiral fermion appears as a pole in the two
point function of an interpolating two-component field $\j$.
It makes no difference whether the interpolating field is elementary
or composite.

  Consider now the self-energy of the interpolating field
$S^{-1}(\o,{\bf p})$. According to our assumption, for small $\o$ and
${\bf p}$, $S^{-1}(\o,{\bf p})\approx \o-\sgmdt{p}$
up to higher order terms. The question is
whether $S^{-1}(\o,{\bf p})$ can have no
other zeros in the Brillouin zone.
Let us consider the function
$\o_0({\bf p})$ defined implicitly by
$\det S^{-1}(\o_0({\bf p}),{\bf p})=0$.
(There are in fact two independent functions which satisfy this
condition because $S^{-1}$ is a two by two matrix). Notice that in
general $\o_0({\bf p})$ will be complex for real values of ${\bf p}$,
because in an interacting theory the self-energy has an imaginary part.

  Suppose that, For small $p$ in the $z$ direction,
$\o_0(0,0,p_3)\approx p_3$. By the periodicity of the Brillouin zone,
$\Re\o_0({\bf p})$ and $\Im\o_0({\bf p})$ should each have another zero
as $p_3$ goes from zero to $2\p/a$.
It now appears that we may be able to avoid an extra zero of
$\o_0({\bf p})$ by going into the complex plane. Namely, by arranging
that $\Im\o_0({\bf p})\ne 0$ at the point where $\Re\o_0({\bf p})=0$.
This, however, is impossible! $\Re\o_0({\bf p})$ is the energy of
the excitation, and an excitation
with vanishing energy cannot have a finite width because of phase space
arguments. Conversly, a finite width would imply that the vacuum is
unstable!

  The above argument indicates that, as long as correlation functions
have ordinary analytical properties, the only new possibility offered
by an interacting theory is that of spontaneous symmetry breaking. In the
present context, the scale set by the expectation value of the would-be
Higgs field tends to be comparable to the
inverse lattice spacing. We can try to fine tune the higgs VEV to a
physical scale, but then the doublers will not decouple because their
masses are proportional to the higgs VEV. For a detailed discussion
of this issue see ref.~[14].

  If we forbid spontaneous symmetry breaking,
we can conclude from the above argument that a doubler must be present.
The purpose of the no-go theorem derived below is to cast this
consideration in a rigorous form.

  We first need some definitions. Given an appropriate set of
interpolating fields, we introduce the {\it retarded anti-commutator}
$$
R_{\a\b}({\bf x},t)=i\th(t)\,\sbra{0}\{\j_\a({\bf x},t)\,,
\j^\dagger_\b({\bf 0},0)\}\sket{0}\,,
\eqno(3)
$$
where possible colour and flavour indices have been suppressed.
We also introduce the space and space-time Fourier transforms
$$
\hat{R}_{\a\b}({\bf p},t)=\sum_{\bf x}
e^{ - i{\bf p \cdot x}}\,R_{\a\b}({\bf x},t) \,,
\eqno(4)
$$
$$
\tilde{R}_{\a\b}({\bf p},\o)=\int_0^\infty dt\, e^{i\o t}
\hat{R}_{\a\b}({\bf p},t)\,,
\eqno(5)
$$
and define
$$
\car_{\a\b}({\bf p})=\lim_{\e\to 0}\tilde{R}_{\a\b}({\bf p},\o=i\e) \,.
\eqno(6)
$$

   On the lattice one cannot demand that (anti)-commutators of
local operators should vanish identically outside the light cone.
We begin in sect.~2 with a discussion of locality properties, which are
classified according to the rate at which $R_{\a\b}({\bf x},t)$
tends to zero at large space-like separations. The main result of
this section is that an exponentially bounded anti-commutator gives
rise to an {\it analytic} $\car({\bf p})$.  The only singularities
occur at {\it generalized degeneracy points}, which are  those
points in the Brillouin zone where the hamiltonian admits
eigenstates of vanishing energy. The proof invokes the
``edge of the wedge'' theorem~[15,16], and it is an adaptation
to the lattice context of classic results from the theory of
dispersion relations.

  In view of the intimate relation between causality and analyticity in
the continuum,  an exponentially bounded anti-commutator on the lattice
should be a necessary condition for causality in the continuum limit.
We comment, however, that in order to apply the NN theorem
all that is needed is that the effective hamiltonian
$\car^{-1}({\bf p})$ have a continuous first derivative. Consequently,
it is sufficient to assume a much weaker form of locality, namely,
that $R_{\a\b}({\bf x},t)$ is bounded by an appropriate inverse power
of ${\bf x}$. This general case will be discussed in sect.~4.
Two examples of anti-commutators in free lattice theories are worked
out in appendix~A.

  In sect.~3 we characterize the possible singularities of
$\car^{-1}({\bf p})$. In view of the desired properties of the
continuum limit (before the gauge interactions are turned on)
we may assume that at sufficiently
large distances physics is correctly described by an effective
lagrangian of massless fermions interaction only via non-renormalizable
couplings. Demanding that all singularities of $\car^{-1}({\bf p})$
should be compatible with those allowed by the effective lagrangian,
we  show that $\car^{-1}({\bf p})$
has regular zeros of the form of eq.~(2) which are in one-to-one correspondence
with massless fermions. Moreover,
$\car^{-1}({\bf p})$ has a continuous first derivative
provided one form of singularity is excluded ``by hand''.
The NN theorem is then applied to $\car^{-1}({\bf p})$
and the above advertized conclusions are obtained.

   The singularity we do not allow is the presence of a {\it pole} in
$\car^{-1}({\bf p})$. We forbid this situation by assuming that the
elements of the matrix $\car_{\a\b}^{-1}({\bf p})$ are bounded.
We comment that, together with the assumption that symmetries
are not broken spontaneously,  this is equivalent via the Ward
identities to the requirement that effective vertex functions, defined
as correlation functions of conserved currents and the interpolating
fields, are pole free.

 The presence of a pole in $\car^{-1}({\bf p})$ may reflect a kinematical
singularity, arising from a bad choice of interpolating fields.
A kinematical singularity can arise, for example,
if one uses two interpolating
fields for two fermions in different corners of the Brillouin zone,
whereas actually they can both be interpolated by a single field,
or if not all fermions are interpolated.
In appendix~B we explain how one can identify the kinematical
nature of the singularity, and describe  a ``trial and error'' method
for constructing an {\it admissible} set of interpolating fields which
is free of kinematical singularities.

  If a pole in $\car_{\a\b}^{-1}({\bf p})$
is not an artifact of an inadmissible set
of interpolating fields, it cannot arise unless the
hamiltonian is highly non-local.
In a very general context, it has been shown that such poles give
rise to the appearance of ghosts in one loop diagrams once gauge
fields are introduced~[8]. The reason is that, via the Ward identity,
such poles appear in the vertex function, but they contribute to
the vacuum polarization with the wrong sign.
Interpreted in a hamiltonian language, this
result implies that the action of a local current on the vacuum takes
one outside the Hilbert space, which is unacceptable. We thus expect
that it should be possible to extend our theorem and to rigorously
exclude the presence of such poles in a consistent quantum theory.

  Sect.~4 is devoted to a generalization of the theorem to a much larger
class of theories. Leaving the other assumptions unchanged, we show
that a sufficient condition for fermion doubling is
that $R_{\a\b}({\bf x},t)$ be bounded  by $1/|{\bf x}|^\g$ where
the exponent $\g$ is strictly bigger than the dimension of space-time.
We believe that this condition is also necessary, but we have not tried
to construct explicit models that can verify this conjecture.

  In sect.~5 we  discuss the implications of our no-go
theorem. The theorem is sufficient to role out the existence of
chiral fermions in all fermion-scalar models proposed so far in the
literature. As already mentioned, this conclusion is in agreement
with explicit model calculations~[11].

  Even more remarkably, the theorem implies that any attempt to
reproduce the standard model on the lattice without violating gauge
invariance must fail, if the
spectrum can be correctly determined by switching off the Electro-Weak
interactions, and the effective vertex functions of the Electro-Weak
currents are pole-free in the symmetric phase.
There is no need to switch off QCD in order to draw
this conclusion! The reason is that the spectrum of QCD does not
contain massless bosons,
and so, in the absence of the photon,
an effective lagrangian of the kind described above
is valid at distances larger than one Fermi. We comment
that it should be possible to accommodated a massless pion without
changing the conclusions because a Goldstone boson has only derivative
couplings.

  The fact that gauge invariant lattice theories are necessarily
vector-like raises an intriguing question concerning the relation between
fermion doubling and the anomaly. If we are careful to work with
an anomaly free theory and to break explicitly at the lattice scale
all global symmetries which are anomalous in the target continuum
theory (\eg baryon number), then there is no ``need''
for the appearance of doublers, because the lattice theory does not have
a bigger symmetry compared to the target theory~[14,17]. This is the
case, for example, in the Eichten-Preskill model~[9]. Nevertheless,
the doublers {\it do} appear. We conclude this paper with a few comments
on possible resolutions of this paradox.

\vspace{5ex}
\noindent {\bf 2.~~Locality and Analyticity}
\vspace{3ex}

  We consider a hamiltonian $\ch$ defined
on a three dimensional regular lattice. The time variable is
continuous, whereas the spacial coordinates take values on a lattice
generated by three linearly independent vectors ${\bf v}_{(k)}$,
$ k=1,2,3$. We assume discrete translation invariance, \ie there exist
three unitary operators $\ct_{(k)}$, commuting among themselves as
well as with the hamiltonian, which generate finite translations
$$
\ct_{(k)}\F({\bf x},t) \ct_{(k)}^\dagger   =
\F({\bf x}+{\bf v}_{(k)},t)\,,
\eqno(7)
$$
where $\F({\bf x},t)$ is a generic field.

  Discrete translation invariance implies the existence of a
Brillouin zone $\cb$. This is the paralleloid spanned by the the three
vectors ${\bf P}_{(j)}$, which satisfy
${\bf P}_{(j)}\cdot{\bf v}_{(k)}=2\p\d_{jk}$. Momentum eigenstates are
labeled by ${\bf p} \in \cb$, and all physical
quantities must be periodic across the Brillouin zone.

  We also assume the existence a set of conserved charges $Q_a$
which are defined as the sum over all lattice points of a local density.
The $Q$-s generate the exact global symmetries of the model.
We assume that they take discrete eigenvalues or, equivalently, that
the global symmetry group is compact. We also assume that the
$Q$-s annihilate the vacuum, \ie there is no spontaneous symmetry
breaking.

  In this paper, we will say that an operator $\co({\bf x},t)$
is local if it depends only on the Heisenberg fields at time $t$ and
at sites ${\bf x'}$ whose distance from ${\bf x}$ is less than a fixed,
finite number. Particularly relevant local operators are
polynomials in the canonical fields, but we can in principle allow
for a more  general class of local operators.

  For a free lattice hamiltonian, relativistic invariance in the low
energy limit amounts to the requirement that all fermionic eigenstates
with vanishing energy compared to the lattice cutoff should be
described by effective two-by-two hamiltonians as in eq.~(2).
But if an interacting theory has massless particles in its
spectrum, it may be difficult (if not impossible)
to separate one particle states from, say,
three particle states containing the original particle plus a soft
fermion-antifermion pair.

  On physical grounds, if the particle can be created in a  causal
process, there should exist a {\it local interpolating field} which
has a finite probability to create the particle by acting on the vacuum.
The particle should then generate a singularity in the two point function
of the interpolating field.

  This consideration suggests the following strategy.
We will not attempt to extract the low energy spectrum from the
elusive one particle states. Instead, we will assume that at sufficiently
large distances, physics is correctly described by an effective
lagrangian of massless fermions interaction only via non-renormalizable
couplings. We will then demand
that all infra-red singularities of the correlation functions of
the lattice interpolating fields should be compatible with those
allowed by the effective lagrangian.

  We will assume that the lattice interpolating fields are local and
that they belong to a given complex representation of the global
symmetry. We intend to study their retarded propagator defined in
eq.~(3), and  to establish its analytical properties.

  We first observe that $R({\bf x},t)$ is {\it bounded}. This is a
trivial consequence of translation invariance and of the fact
that $\j_\a({\bf x},0)$ is a well defined operator on
the Hilbert space. Thus, there exists $0<b_1<\infty$ such that
$$
|R_{\a\b}({\bf x},t)| \le
\left\Vert
\stackrel{ }{\j({\bf 0},0)}\!
\sket{0}\right\Vert^2 +
\left\Vert
\j^\dagger({\bf 0},0)\sket{0}\right\Vert^2
\le b_1 \,.
\eqno(8)
$$

  Another important property is that for fixed ${\bf x}$,
$R({\bf x},t)$ is an analytic function of $t$. Once more, this is a
trivial consequence of the fact that on the lattice all operators
are well defined and that
$\j({\bf x},t)=\exp(i\ch t) \j({\bf x},0) \exp(-i\ch t)$.
As a result, $R({\bf x},t)$ cannot vanish identically outside the light
cone, for then it would be zero everywhere.

  We should therefore distinguish between theories with various degree of
locality, as characterized by the rate at
which $R({\bf x},t)$ tends to zero at large space-like separations.
We will say that $R({\bf x},t)$ is {\it weakly local of degree} $\g>0$
if there are positive constants $c$ and $b_2$
such that for all $|{\bf x}|>ct$
$$
|R_{\a\b}({\bf x},t)| \le
\min \left\{b_1, {b_2\over (\,|{\bf x}|-ct)^{\g}}\right\} \,.
\eqno(9a)
$$
Notice that this is an inverse power law bound.

  We will say that $R({\bf x},t)$ is {\it local} if it can be bounded
by an exponential, \ie if there are positive constants $c$, $b_2$
and $\m$ such that for all $|{\bf x}|>ct$
$$
|R_{\a\b}({\bf x},t)| \le
\min \left\{b_1, b_2 e^{-\m(|{\bf x}|-ct)}\right\} \,.
\eqno(9b)
$$
Finally, we will say that $R({\bf x},t)$ is {\it strongly local} if it
decreases
faster than an exponential, \ie if it satisfies a bound of
the form~(9b) for every $\m$.

  In this section we will assume that $R({\bf x},t)$ is bounded by an
exponential. This will allow us to prove that $\car({\bf p})$ is
analytic. The more general case will be treated in sect.~4.
As suggested by the example of appendix~A, it is plausible
that short range lattice hamiltonians always give rise to strongly local
(anti)-commutators. In any event, as we have already mentioned, an
analytic propagator is compatible with causality in the continuum limit.
We therefore expect that only theories with exponentially bounded
(anti)-commutators can have a consistent continuum limit. It is
interesting, though, that fermion doubling can be proved assuming only
that $R({\bf x},t)$ is weakly local
of degree $\g>d$ ($d$ is the space-time dimension). In this case
$\car({\bf p})$ will have a continuous first derivative except at
generalized degeneracy points. As shown in appendix~A,  the
anti-commutator of SLAC fermions violates this condition,
as it must do in order to allow for the discontinuity of the
spectrum at $p=\p/a$.

  The last thing we need is the notion of a {\it generalized degeneracy
point}. We introducing the {\it advanced} anti-commutator
$$
A({\bf x},t)=-i\th(-t)\,\sbra{0}\{\j_\a({\bf x},t)\,,
\j^\dagger_\b({\bf 0},0)\}\sket{0}\,,
\eqno(10)
$$
as well as its Fourier transform $\tilde{A}({\bf p},\o)$.
For real values of ${\bf p}$ and $\o$, we define
$$
E_0({\bf p})=\sup\{\o\,|\,\tilde{R}({\bf p},\o')=
\tilde{A}({\bf p},\o')\mbox{  if  } |\o'|<\o\}\,.
\eqno(11)
$$
The physical meaning of this definition is that $E_0({\bf p})$ is the
lowest possible energy for eigenstates with momentum ${\bf p}$.
We define a generalized degeneracy point by the condition
$E_0({\bf p}_c)= 0$. Thus, ${\bf p}_c$ is a generalized degeneracy point
if it is the end point of a gap-less continuous spectrum.

  We now establish the basic analytical properties of the retarded
propagator.

\vspace{2ex}

\noindent {\it Lemma}. Assume that $R(x,t)$ is local in the sense
of eq.~(9b). Then (a) $\tilde{R}({\bf p},\o)$ is holomorphic in the
domain $\Im\o>0$, $|\Im{\bf p}|<\min\{c^{-1}\Im\o\,,\m\}$; (b)
$\car({\bf p})$ is analytic with singularities only at generalized
degeneracy points.

\vspace{1ex}

\noindent {\it Proof}. By assumption, the \rhs of eq.~(4) is bounded
by the \rhs of eq.~(9b) times $e^{|{\bf x}|\,|\Im{\bf p}|}$.
Hence, the sum in eq.~(4) converges absolutely and
$\hat{R}_{\a\b}({\bf p},t)$ is holomorphic in the domain
$|\Im{\bf p}|<\m$.
Morever, $\hat{R}_{\a\b}({\bf p},t)$ can be bounded by a polynomial of
third degree in $t$ times $e^{ct\,|\Im{\bf p}|}$.
The presence of the damping factor
$e^{-t\,\Im \o}$ then implies that the integral on the \rhs of eq.~(5)
converges absolutely for $|\Im{\bf p}|<\min\{c^{-1}\Im\o\,,\m\}$.
This proves (a).  Notice that if  $R(x,t)$ is strongly local
than $\hat{R}_{\a\b}({\bf p},t)$ is an entire function of ${\bf p}$.
In this case $\tilde{R}({\bf p},\o)$ is holomorphic in the forward cone
$|\Im{\bf p}|<c^{-1}\Im\o$.

 In order to prove (b) we notice that the Fourier transform of the
{\it advanced} anti-commutator has similar properties except that the
sign of $\Im\o$ is now negative.
A straightforward application of the edge of the wedge theorem~[15]
now implies that the common boundary function $\car({\bf p})$ is
analytic, with singularities only at generalized degeneracy points.
(A modern proof of the theorem can be found for example in ref.~[16]).
This proves~(b).

\vspace{1ex}

  The analyticity of $\car_{\a\b}({\bf p})$ away from generalized
degeneracy points, implies that there can be no obstructions to the
smooth motion throughout the Brillouin zone from one zero of
$\car_{\a\b}^{-1}({\bf p})$ to another,  provided we exclude the
possibility of {\it poles} in  $\car_{\a\b}^{-1}({\bf p})$.
We forbid this situation by assuming that the elements of the matrix
$\car_{\a\b}^{-1}({\bf p})$ are bounded. The justification for
this assumption has been discussed in the introduction. As we have
explained there, we believe that this is a pathological situation, and
that in a more complete treatment it should be possible to exclude it
rigorously. Finally, we remind the reader that this assumption implies in
particular the absence of kinematical singularities and hence that the
set of interpolating fields is admissible. For more details see
appendix~B.

\vspace{5ex}
\noindent {\bf 3.~~A no-go theorem}
\vspace{3ex}

  By definition, $\car^{-1}({\bf p})$ is a hermitian matrix.
In order to show that it qualifies as an effective hamiltonian which
satisfies the NN theorem, what is left for us to do is to
show that it has a continuous first derivative at
generalized degeneracy points and that its zeros can be
identified with massless fermions.

  We assume that the continuum limit is relativistic and that
the only massless particles are fermions.
Low energy physics should therefore be described by an effective
lagrangian containing massless fermions coupled only via
non-renormalizable interactions. The interaction of the smallest
possible dimension is a four Fermi coupling.
Notice that the masses of other excitations need not
diverge as the lattice spacing tends to zero. There may well be massive
excitations in the {\it physical} spectrum. In this case the above
effective lagrangian correctly describes physics at distance scales
which are sufficiently large compared to the Compton wave length of
the least massive particle.

  Let us denote by ${\bf p}_{phys}$ the momentum variable which
transforms homogeneously under the Lorentz group in the low energy limit.
The only form of singularity in the zero frequency fermionic propagator
which is compatible with the above effective lagrangian is
$$
\pm{1\over \sgmdt{p}_{phys}
\left(1+O({\bf p}_{phys}^4 \log{\bf p}_{phys}^2)\right) }\,.
\eqno(12)
$$
We have ${\bf p}_{phys}^4$ in front of the logarithmic term
because this term involves at least two powers of coupling constants,
and all coupling constants have a negative mass dimension which is
at least two.

  Clearly, an allowed singularity of $\car({\bf p})$ can be obtained by
substituting ${\bf p - p}_c$ instead of ${\bf p}_{phys}$ is eq.~(12).
We call such a singularity a {\it primary singularity}. The precise definition
is as follows. A generalized degeneracy point ${\bf p}_c$
is a primary singularity if there exists a unitary transformation $U$
such that
$$
\lim_{ {\bf p}\to{\bf p}_c}
\left[ \sgmdt{(p-p}_c)\otimes I_1 \right]
U \car({\bf p}) U^{-1} = \left[ I \otimes A \right]\,.
\eqno(13)
$$
In eq.~(13), $I$ is the identity matrix in spin space, $I_1$ is the
identity matrix in colour and flavour space, and $A$ is a diagonal
matrix $A=diag(Z_1,\ldots,Z_k,0,\ldots,0)$. The $Z$-s are non-zero
constants, which are in one-to-one correspondence with massless
fermions. The helicity of the massless fermion is determined by the
sign of the corresponding $Z$. We comment that if one does not insists
that the massless fermion will propagate at the speed of light, than one
can allow for any non-singular linear transformation instead of $U$.
But the relativistic case is obviously more restrictive.

  We now have to determine whether there are other allowed forms of
singularity points. In order to do so we have to digress for a
moment and discuss what are the acceptable values of ${\bf p}_c$
for a primary singularity point. We claim that Poincar\'{e} invariance
in the continuum limit implies that a primary singularity can occur
only at a point which is a linear combination with {\it rational}
coefficients of the three vectors ${\bf P}_{(j)}$ which span the
Brillouin zone.

  Assuming that this property holds and that the number of
primary singularity points is finite (\ie a finite number of massless
fermions), there exist three smallest positive integers $n_{(j)}$
such that all primary singularity points are representatives of the
origin of a reduced Brillouin zone $\cb'$ spanned by
${\bf P}_{(j)}/n_{(j)}$. We can
then identify the physical momentum with ${\bf p}\in\cb'$.

  Let us now consider what other
forms of singularities are allowed in $\car({\bf p})$. As an example,
take a cubic lattice and assume that a primary singularity occurs
at ${\bf p}_c=(2\p/Na,0,0)$ for some $N$. The presence of a single
particle spectrum near ${\bf p}_c$ then implies the existence of
multi-particle spectra at the points $n{\bf p}_c$ for $n=2,\ldots,N$.
The multi-particle thresholds will all be at $E=0$ because our
particle is massless. This implies that all these points are generalized
degeneracy points of the hamiltonian.

 In general, the quantum numbers
of the multi-particle states may be different from those of the
original particle. But since the global symmetry group is compact,
some of these states {\it will} have the same quantum numbers as the
original particle and so they will contribute to $\car({\bf p})$
if the are no additional accidental symmetries.
We will denote the corresponding points as secondary singularities.
The leading contribution of the gap-less spectrum to
$\car({\bf p})$ at a secondary singularity point can take the form
$$
\pm \sgmdt{p}_{phys}\,{\bf p}_{phys}^2 \log{\bf p}_{phys}^2\,.
\eqno(14)
$$
As before, ${\bf p}_{phys}={\bf p}-{\bf p}_c$.
This form is dictated by the requirement that, once we sum
$\car({\bf p})$ over all points that correspond to the same
${\bf p}_{phys}$, we will obtain an expression compatible with the
expansion of the denominator of eq.~(12) in powers of the coupling
constant. Notice that there is no reason that $\car({\bf p})$ should
vanish at a secondary singularity point, because it always receives
additional, regular contributions from finite energy branches of the
spectrum.

  In ${\bf x}$-space, the requirement that all singularity points
occur at rational points means that zero energy states are periodic.
Eigenstates of definite physical momentum are then obtained by a
Fourier sum over sub-lattices which respects that periodicity.
Zero energy states are constant on every sub-lattice, and this is
the lattice equivalent of the property that a relativistic continuum
eigenstate of zero energy must be constant in space.

  By contrast, suppose that a primary singularity occurred at
a non-rational point. This would imply the the lattice model has a
non-periodic zero energy state. Moreover, this primary singularity
would lead to an infinite number of secondary singularities that
would be dense on a sub-space of the Brillouin zone whose dimension
is at least one. Under these circumstances the continuum limit cannot
be both translation invariant and Lorentz invariant.
It would be impossible to define a conserved momentum  which generates
translations in space, has a continuous spectrum and which, at the same time,
assigns the eigenvalue zero to all zero energy states.

  In fact, under these circumstances,
insisting that a ``momentum'' operator assign the eigenvalue
zero to all zero energy states would entail the appearance of infinitely
many new conservation laws\footnote{
This observation is due to A.~Casher.}. Two particles whose total
``momentum'' is zero would in general never scatter into a final state
with the same ``momentum'' and the same quantum numbers under the
lattice internal symmetries, because in general the initial and
final states would have different lattice momenta. Moreover, the number
of superselection rules not explained by the lattice internal symmetries
and conservation of this ``momentum'' would be infinite.

  We now collect all our intermediate results together in the following
theorem.

\vspace{2ex}

\noindent {\it Theorem}. Consider a hamiltonian defined on a regular
lattice. Assume the existence of a compact global symmetry group which
is not spontaneously broken. Assume also that the continuum limit is
relativistic and that the only massless particles are fermions.
Under these assumptions, there is an equal number of left handed and
right handed fermions in every complex representation of the global
symmetry group, provided $R({\bf x},t)$ is local and
$\car^{-1}({\bf p})$ is bounded for every admissible set of
interpolating fields.

\vspace{1ex}

\noindent {\it Proof}. Consider all sets of interpolating fields which
satisfy the above assumptions and which belong to a given complex
representation. Choose a maximal set. By this we mean that the total
number of $Z$-s, as determined by the limiting procedure~(13) and summed
over all primary singularity points, is maximal. This number is then
the total number of massless fermions in that representation.

  Locality of $R({\bf x},t)$ and boundedness of $\car^{-1}({\bf p})$
imply that $\car^{-1}({\bf p})$ is analytic except at generalized
degeneracy points. Furthermore, the allowed forms of singularities,
eqs.~(12) and~(14), imply that $\car^{-1}({\bf p})$ has a continuous
first derivative at the generalized degeneracy points, and that all
zeros of $\car^{-1}({\bf p})$ (which occur at primary singularity
points) are of the relativistic form~(2). In addition,
$\car^{-1}({\bf p})$ is hermitian. Hence $\car^{-1}({\bf p})$ satisfies
all the assumptions of the NN theorem. Applying the
theorem, we conclude that $\car^{-1}({\bf p})$ has an equal number of
left handed and right handed zeros. Since the set of interpolating field
we have chosen is maximal, this implies that the spectrum contains
an equal number of left handed and right handed fermions in the given
complex representation.

\vspace{1ex}

  For completeness, we recall why the assumption that a fermion belongs
to a complex representation is needed in the NN theorem.
If a fermion belongs to a real representation, it is possible to use
real field (Majorana) formulation. A single Majorana fermion can then
generate both a left handed and a right handed pole in its two point
function. This is the only way to violate the one-to-one correspondence
between poles of the two point function and massless fermions. Of course,
this exceptional situation is of no help if we are trying to construct
chiral fermions.

\vspace{5ex}
\noindent {\bf 4.~~The most general case}
\vspace{3ex}

  We now turn to discuss the most general case. In this section we will
prove that the spectrum is vector-like provided $R({\bf x},t)$ is weakly
local of degree $\g>4$, and all other assumptions are the same as in
sect.~3. (For simplicity we continue to work in four dimensions).
We believe that this condition is not only sufficient, but actually
necessary. However, we have not attempted to construct explicit models
that will assess this conjecture.

  In complete analogy to the first part of the lemma of sect.~2, one can
show that if $R({\bf x},t)$ is weakly local of degree $\g>3$ than
$\hat{R}_{\a\b}({\bf p},t)$ is continuous. Moreover, since taking
the $n$-th ${\bf p}$-derivative amounts to multiplication by
${\bf x}^n$ in configuration space, $\hat{R}_{\a\b}({\bf p},t)$ will
have continuous $n$-th derivatives with respect to ${\bf p}$ provided
$R({\bf x},t)$ is weakly local of degree $\g>3+n$. Since we need a
continuous first derivative, we assume that $R({\bf x},t)$ is weakly
local of degree $\g>4$.

  What is left for us to do is to prove that, if
$\hat{R}_{\a\b}({\bf p},t)$ has continuous $n$-th derivatives with
respect to ${\bf p}$, then the same is true for $\car({\bf p})$
except at generalized degeneracy points. The assumed form of the
low energy effective lagrangian is the same as before, and so from this
point on we can simply repeat the discussion of sect.~3.

  In order to relate the differential properties of
$\hat{R}_{\a\b}({\bf p},t)$ and $\car({\bf p})$ we introduce the
spectral representation
$$
\tilde{R}_{\a\b}({\bf p},\o)= \int_0^{\infty}dE\,
\left\{ { \r^{(1)}_{\a\b}({\bf p},E) \over E-\o } -
        { \r^{(2)}_{\a\b}({\bf p},E) \over E+\o } \right\}\,.
\eqno(15)
$$
In eq.~(15) $E$ and ${\bf p}$ are real, whereas $\o$ is complex with
$\Im\o>0$. As before, we are interested in the limit $\Im\o\to 0$.
The spectral functions are given by
$$
\r^{(1)}_{\a\b}({\bf p},E) = \int_n \d(E-E_n) \d^3({\bf p}-{\bf p}_n)
\sbra{0}\j_\a({\bf 0},0)\sket{n}
\sbra{n}\j^\dagger_\b({\bf 0},0)\sket{0}\,,
\eqno(16)
$$
and $\r^{(2)}_{\a\b}({\bf p},E)$ is obtained by interchanging $\j_\a$
and $\j^\dagger_\b$ and replacing $\d^3({\bf p}-{\bf p}_n)$ by
$\d^3({\bf p}+{\bf p}_n)$. In eq.~(16) $\int_n$ stands for a sum over
all intermediate states. One can relate the two spectral functions
by invoking $\cp\cc\ct$ invariance, but we will not need this relation
below.

  We will first prove the continuity of $\car({\bf p})$ away from
generalized degeneracy points. Let us denote
$$
\O^{(i)}_{\a\b}({\bf p})= \int_0^{\infty}dE\,
\r^{(i)}_{\a\b}({\bf p},E)\,,\quad\quad i=1,2\,.
\eqno(17)
$$
We want to prove the existence of a uniform bound
$$
|\O^{(i)}_{\a\b}({\bf p})|<b_3\,,
\eqno(18)
$$
for some $0<b_3<\infty$. Once this bound is established we are done.
While we know very little on $E_0({\bf p})$ for general values of
${\bf p}$, our assumption that low energy physics is relativistic
implies in particular that $E_0({\bf p})$
is continuous provided its value is less then some physical scale
$E_1>0$. Therefore, for every  ${\bf p}\in\cd$, where $\cd$ is a small
enough neighbourhood of ${\bf p}_0\in\cb$,  one has
$$
 E_m({\bf p}_0) \le E_0({\bf p})\,,
\eqno(19)
$$
$$
E_m({\bf p}_0)=\min\{ \raisebox{.25ex}{${\scriptstyle \half}$}
E_0({\bf p}_0)\,,E_1\}\,.
\eqno(20)
$$
Like $E_0({\bf p}_0)$, one has $E_m({\bf p}_0)=0$ if and only if
${\bf p}_0$ is a generalized degeneracy point. Assuming that ${\bf p}_0$
is {\it not} a generalized degeneracy point,
it is now straightforward to establish the following uniform
bound for every ${\bf p}\in\cd$
\beqr{21}
|\tilde{R}_{\a\b}({\bf p},i\e) & \! -
& \! \tilde{R}_{\a\b}({\bf p},i\e')| \le \NON
& \le & \int_{E_0({\bf p})}^{\infty}dE\, \left(
\left|\r^{(1)}_{\a\b}({\bf p},E)\right|+
\left|\r^{(2)}_{\a\b}({\bf p},E)\right|\right)\,
{ |\e-\e'| \over E_0^2({\bf p}) } \NON
& \le & { 2b_3\over  E_m^2({\bf p}_0) } |\e-\e'| \,.
\eeqr
Eq.~(21) implies that $\car_{\a\b}({\bf p})$  is continuous except at
generalized degeneracy points.

  It remains to establish the bound~(18). Since the condition
of weak locality holds for every $t$, it follows that the magnitude of
long range couplings in the hamiltonian should obey the same bound~(9a)
as $R_{\a\b}({\bf x},t)$, where we now set $t=0$.
In quantum mechanics, time derivatives are given
by multi-commutators with the hamiltonian, and so
the $n$-th time derivative of $R_{\a\b}({\bf x},t)$ obeys the same
bound at large space-like separations as $R_{\a\b}({\bf x},t)$ itself.
For the case at hand, $\g>3$, the ${\bf x}$-summation converges
uniformly for every $n$, and so
$\hat{R}_{\a\b}({\bf p},t)$ has continuous time derivatives of every
order.

The spectral representation for the $n$-th time derivative is
$$
\left. \left( i{d\over dt}\right)^{(n)}
\hat{R}_{\a\b}({\bf p},t)\,\right|_{t=0}=
\int_0^{\infty}dE\, E^n
\left\{ \r^{(1)}_{\a\b}({\bf p},E) + (-)^n
        \r^{(2)}_{\a\b}({\bf p},E) \right\}\,.
\eqno(22)
$$
Convergence of the integral on the \rhs of eq.~(22) for even $n$ implies
that $|\r^{(1)}_{\a\b}({\bf p},E) + \r^{(2)}_{\a\b}({\bf p},E)|$
decreases faster than any power of $E$ for $E\to\infty$. Convergence
of the integral for odd $n$ implies the same for the difference of the
spectral functions. Hence each of the spectral functions separately
decreases faster than any power. Together with continuity of
$\hat{R}_{\a\b}({\bf p},t)$ as a function of ${\bf p}$, this implies
that $\O^{(i)}_{\a\b}({\bf p})$ as defined in eq.~(17) is continuous.
Compactness of the Brillouin zone than implies the existence of the
uniform bound~(18). This concludes the proof that $\car({\bf p})$
is continuous except at generalized degeneracy points.

  The proof that the derivative of $\car({\bf p})$ is continuous
except at generalized degeneracy points
is completely analogous. We simply replace every function of ${\bf p}$
by its ${\bf p}$-derivative in eqs.~(15-22). In general, this process
can continue up to the $n$-th ${\bf p}$-derivative,
provided $R_{\a\b}({\bf x},t)$ is weakly local of degree $\g>3+n$.

  In conclusion, we showed that if $R_{\a\b}({\bf x},t)$
is weakly local of degree $\g>4$ than $\car({\bf p})$ has a
continuous first derivative
except at generalized degeneracy points. From here on one can repeat
the discussion of sect.~3 and prove that the spectrum is necessarily
vector-like.

\vspace{5ex}
\noindent {\bf 5.~~Discussion}
\vspace{3ex}

  The main lesson of our no-go theorem is that, as far as fermion
doubling is concerned, there is essentially no difference between free
and interacting lattice theories. From the correlation functions of the
interacting theory one can always construct an object which plays the
role of an effective quadratic hamiltonian for the massless fermions.
The types of singularities in the dispersion relation which allow one
to escape the conclusions of the no-go theorem are the same as in the
free case. Thus, the only remaining question is whether such ``bad''
singularities can be less dangerous if they occur in an interacting
theory. In this paper we have gone a
long way towards giving a rigorous, negative answer to this question.
For example, we can reject without relying on
perturbation theory~[18] any lattice model whose
dispersion relation for the effective low energy degrees of freedom
contains a discontinuity, on the grounds that such a theory must be
a-causal.

  Our no-go theorem is sufficiently strong to exclude the existence of
chiral fermions in all the fermion-scalar models proposed for that
purpose in the literature~[9]. This is no surprise, as all explicit
model calculations~[11] have consistently reached the same conclusion.
The virtue
of our approach is that it provides a uniform basis for treating all
these models, and that it constraints the spectrum by invoking only
physical properties which characterizes a consistent continuum limit.

  As we have already mentioned in the introduction, perhaps the most
striking consequence of our theorem is the
constraints it puts on any attempt to
reproduce the standard model on the lattice without violating gauge
invariance. It asserts that any such attempt must fail, if the
spectrum can be correctly determined by switching off the Electro-Weak
interactions (there is no need to switch off QCD), and provided that the
effective vertex functions of
the Electro-Weak currents are pole-free in the symmetric phase.

  One may question the relevance of this conclusion on the grounds that
we are not dealing directly with {\it gauged} Electro-Weak interactions.
But when one says that the a gauge theory is chiral, one is making a
statement on its elementary, gauge variant fields!
These cannot be related directly to physical observables
without invoking additional assumptions such as complementarity.
Thus, the very
${\it definition}$ of a chiral gauge theory implicitly assumes that
we can determine the elementary fermions' content by setting the
gauge coupling to zero.

  Moreover, the widely accepted view that QCD, as defined on the
lattice, is a consistent quantum theory with a non-trivial continuum
limit, is based on asymptotic freedom and the validity of weak coupling
perturbation theory at short distances. Thus, in our opinion, one cannot
reject our conclusions without, at the same time, questioning the
validity of the present understanding of non-abelian gauge theories.

  From a conceptual point of view, this situation is very intriguing,
for Electro-Weak phenomenology has been found to be in remarkable
agreement with the predictions of continuum perturbation theory, and
the lattice has been introduced mainly to allow for a quantitative
treatment of the strong interactions. Here we are faced with an
inability to provide a uniform, consistent treatment to all the
interactions of the standard model simultaneously.

  Paradoxically, the only approach to putting the weak interactions
on the lattice which still seems viable is based on using gauge
non-invariant actions. Several such proposals exist in the
literature~[19].
We include among them the Rome approach~[20], because one cannot derive
the gauge fixed action of this approach from a gauge invariant
lattice action. It has also been claimed~[21] that the recently proposed
domain wall fermions do not give rise to a chiral spectrum unless one
gives up tree level gauge invariance.

  So far, there are only partial results concerning these models,
and all of them are in the context of
perturbation theory. At this level,
it is not unreasonable to expect that gauge invariance will
be recovered in the continuum limit, provided the spectrum is anomaly
free. In this respect the lattice is not very different from continuum
field theories, where no gauge invariant regularization for chiral
theories is known to exist. Nevertheless, the Adler-Bardeen theorem~[22]
gauranties the consistency of perturbation theory provided the theory
is anomaly free at the one loop level. In modern language, the
Adler-Bardeen theorem can be derived using renormalization group
arguments~[23]. The renormalization group exists on the lattice as well,
and so it is not unreasonable to expect that some version of the
Adler-Bardeen theorem should hold on the lattice.

  The real unknown is the behaviour of such models at the
non-perturbative level. By the no-go theorems, gauge invariance
on the lattice implies a vector-like spectrum if we demand that
the continuum limit be free of inconsistencies such as violations
of causality or Lorentz invariance. Thus, one can suspect that models
with explicitly broken gauge symmetries will reveal some unexpected
features at the non-perturbative level. Until a detailed understanding
of non-perturbative effects in such models is reached, their consistency
as well as their relevance to the real world remain unclear.

\vspace{5ex}
\centerline{\bf Acknowledgements}
\vspace{3ex}

  I thank A.~Casher for numerous discussions of the subject.
This research was supported in part by the Basic Research Foundation
administered by the Israel Academy of Sciences and Humanities, and by
a grant from the United States -- Israel Binational Science Foundation.

\vspace{5ex}
\noindent {\bf Appendix~A}
\vspace{3ex}

  In this appendix we give two examples of anti-commutator functions
in free fermionic hamiltonians in one space dimension.
The retarded anti-commutator is obtained by simply multiplying
the ordinary anti-commutator by $\th(t)$.

 Let us define
$$
\D_F(x,t)=\bra{0}\{\j(x,t)\,, \j^\dagger(0,0)\}\ket{0}\,.
\eqno(A.1)
$$
Here $\j(x,t)$ is a single component fermion field. In this appendix and
the next one we set $a=1$, and the space coordinate $x$ takes integer
values. $\D_F(x,t)$ satisfies the homogeneous wave equation
$$
(\partial_t - \nabla ) \D_F(x,t) =0 \,,
\eqno(A.2)
$$
where $\nabla$ is a lattice difference operator, and the boundary
condition $\D_F(x,0)=\d_{x,0}$. It is convenient to express
$\D_F(x,t)$ as
$$
\D_F(x,t)=-i(\partial_t + \nabla ) \D(x,t) \,,
\eqno(A.3)
$$
where $\D(x,t)$ is a homogeneous solution of the second order equation
$$
(\partial_t^2 - \nabla^2 ) \D(x,t)=0 \,,
\eqno(A.4)
$$
with the boundary conditions
$\D(x,0)=0$ and $\partial_t\D(x,0)=i\d_{x,0}$.
An explicit representation for $\D(x,t)$ is
\beqrabc{A.5}
\D(x,t) & = & {1\over 2\p}\Int\o\dInt{-\p}{\p}{p} e^{i\o t-ipx}
\d(\o^2-E^2(p))\,\e(\o) \\
& = & {1\over 2\p}\int_{-\p}^{\p}{dp\over 2|E(p)|}\,
\left(e^{i|E(p)|t-ipx}-e^{-i|E(p)|t-ipx}\right)\,.
\eeqr
Here $E(p)$ is the dispersion relation as determined by the free
hamiltonian.

  Consider first a nearest neighbour hamiltonian. The lattice
difference operator is $\nabla= \half(\d_{x,x'+1}-\d_{x,x'-1})$, and
the dispersion relation is $E(p)=\sin p$. It is convenient to calculate
$\partial_t\D(x,t)$ first. A straightforward calculation gives rise to
$$
\partial_t\D(x,t)=\left\{ \begin{array}{ll}
0\,, & x=2n+1 \\ iJ_{|x|}(t)\,, & x=2n\,.
\end{array} \right.
\eqno(A.6)
$$
Using eq.~(A.3) and the boundary conditions satisfied by $\D(x,t)$
we obtain
$$
\D_F(x,t)=\left\{ \begin{array}{ll}
J_{|x|}(t)\,, & x\ge 0\,, \\ (-)^x J_{|x|}(t)\,, & x\le 0\,.
\end{array} \right.
\eqno(A.7)
$$
This result clearly exhibits the presence of a primary right mover and
a doubler which is a left mover. We can estimate $\D_F(x,t)$ using
the asymptotic expansion for Bessel functions where both the order
$|x|$ and the argument $t$ are large. Denote $\x=(|x|-t)/t$.
Assuming $\x\ll 1$ and $t\x\gg 1$ one has
$$
J_{|x|}(t)\approx\left\{ \begin{array}{ll}
{\exp(-t\sqrt{2\x^3}) \over t\sqrt{2\x} }\,, & \x>0\,, \\
{\cos(t\sqrt{-2\x^3}) \over t\sqrt{-2\x} }\,, & \x<0\,.
\end{array} \right.
\eqno(A.8)
$$
The slow decay of $\D_F(x,t)$ {\it inside} the light cone signals the
presence the the high energy excitations with group velocity which is
smaller than one. Outside the light cone, $\D_F(x,t)$ decreases faster
than an exponential.

  We now perform a similar calculation for the SLAC derivative.
The dispersion relation is $E(p)=p$. We now have
\beqr{A.9}
\D_F(x,t) & = & {1\over 2\p}\int d\o\int_{-\p}^{\p}dp\,
(\o+p) e^{i\o t-ipx}
\d(\o^2-E^2(p))\e(\o) \NON
& = & {1\over\p} {\sin(\p(x-t))\over x-t} \,.
\eeqr
Thus, although $\D(x,t)$ is maximal on the classical trajectory $x=t$,
it decays extremely slowly for both $|x|<t$ and $|x|>t$. The non-local
character of the anti-commutator is necessary in order to produce
the expected discontinuity in $\car(p)$.

\vspace{5ex}
\noindent {\bf Appendix~B}
\vspace{3ex}

  An unwise choice of interpolating fields may give rise to
two types of spurious, kinematical singularities in $\car^{-1}({\bf p})$.
This can best be illustrated
through an example in the context of a free fermion theory. Consider
again a single component field in one space dimension with a
standard nearest neighbour hamiltonian. The spectrum consists of a
right mover at $p=0$, and a left mover at $p=\p$.
In self-explanatory notation, the retarded
anti-commutator is
\beqr{B.1}
\car(p) & = & \svev{\j\,\j^\dagger}(p) \NON
& = & {1\over\sin p}\,.
\eeqr

  Suppose that, instead of using $\j(x)$ as a single interpolating
field for both fermions,
we unwisely decide to use the two fields
$\j_{\pm}(x)=\pm\half(\j(x+1)+\j(x-1)\pm 2\j(x))$. In Fourier space this
becomes $\j_{\pm}(p)=(1 \pm \cos p) \j(p)$. The $\pm$ is now
considered as a ``flavour'' index, and
$$
R'(x,t)=\left( \begin{array}{cc}
\svev{\j_+\j_+^\dagger} & \svev{\j_+\j_-^\dagger} \\
 & \\
\svev{\j_-\j_+^\dagger} & \svev{\j_-\j_-^\dagger}
\end{array} \right)\,.
\eqno(B.2)
$$
In Fourier space this becomes
$$
\car'(p)=\left( \begin{array}{cc}
(1+\cos p)^2 & \sin^2 p \\
\sin^2 p & (1-\cos p)^2
\end{array} \right) {1\over\sin p}\,.
\eqno(B.3)
$$
Notice that ${1\over\sin p}$ is nothing but $\car(p)$. We now observe
that $\det \car'(p)=0$ identically. This is a consequence of the fact
the the two fields $\j_{\pm}(x)$ are {\it linearly dependent in the
Hilbert space}. By this we mean that their matrix elements between
any two states of momentum $k$ and $k+p$ are linearly dependent for
fixed $p$. In the above example, the proportionality constant is
the kinematical factor $(1+\cos p)/(1-\cos p)$.

  The solution to this problem is simple. In the above
example, we have to replace the two fields $\j_{\pm}(x)$ by a linear
combination of them $a\j_+(x)+b\j_-(x)$. In so doing, there are some
constraints on the coefficients $a$ and $b$ which are necessary in order
to prevent the appearance of spurious poles in $\car({\bf p})$.

  Clearly, such a pole will appear if
one of the coefficients is zero. In this case we are interpolating only
one of the fermions, and a {\it zero in the propagator} will appear at
the location of the missing fermion. Another possibility is that we
take both $a$ and $b$ to be non-zero, but with opposite signs. What is
common to both is the existence of a point where the linear combination
$a(1+\cos p)+b(1-\cos p)$ vanishes. On the other hand, any choice of
$a$ and $b$ where both are, say, strictly positive, is admissible.
In this case the kinematical factor never vanishes and both fermions
are interpolated. Notice that the choice $a=b=1/2$ brings us back to
the original field $\j(x)$.

  Of course,  similar phenomena can arise from a bad choice of
interpolating  fields in an interacting theory as well. What we need is
a procedure that will allow us eliminate all kinematical singularities.
Since we have no knowledge on the way a given set of interpolating fields
was constructed, we are lead to the following two step procedure.

  In the first step we {\it on purpose} multiply the interpolating fields
by various kinematical factors. If necessary, we enlarged the set by
additional fields that are linearly dependent in the Hilbert space on
existing ones. The aim at this stage is to achieve a one-to-one
correspondence between fields and fermions.

  In the second stage we minimize the total number of interpolating fields
by taking appropriate linear combinations of the previous, linearly
dependent fields.
Suppose that there are $n$ primary singularities, and $k_n$
fermions are interpolated at the $n$-th primary singularity. We can than
replace the fields which interpolate the first fermion at  every point
by a single linear combination. This single field will interpolate
the first fermion at every primary singularity.

  If we continue this process, we are guaranteed that the final set
will be independent in the Hilbert space. However, without additional
information, we do not know whether in every step it will be possible
to avoid the appearance of zeros in the propagator. If no choice of
interpolating field can be made which is free of such zeros, than these
zeros are genuine, and the pole in the inverse propagator is not a
kinematical singularity. As we have discussed in the introduction, we
expect that theories with this property must be inconsistent.

  Finally, we comment that the complicated process described above was
necessary only because we assumed that we know nothing about the
hamiltonian of the theory, nor on the way the interpolating fields are
constructed from the elementary fields of the theory. In practise, one
has a specific model in mind, and it is easy to identify  reasonable
candidate interpolating fields as well as to verify that no kinematical
singularities are present.

\vspace{5ex}
\centerline{\rule{5cm}{.3mm}}

\vspace{3ex}
\newcounter{00001}
\begin{list}
{[~\arabic{00001}~]}{\usecounter{00001}
\labelwidth=1cm}

\item K.G.~Wilson, in {\it New Phenomena in Sub-Nuclear Physics} (Erice, 1975),
ed. A.~Zichichi (Plenum, New York, 1977).

\item J.~Smit, unpublished.

\item L.H.~Karsten and J.~Smit, \NPB{183} (1981) 103.

\item H.B.~Nielsen and M.~Ninomiya, \NPB{185} (1981) 20,
{\it Errata} \NPB{195} (1982) 541; \NPB{193} (1981) 173.

\item S.D.~Drell, M.~Weinstein and S.~Yankielowicz, \PRD{14} (1976) 487, 1627.

\item C.~Rebbi, \PLB{186} (1987) 200.

\item L.H.~Karsten and J.~Smit, \NPB{144} (1978) 536; \PLB{85} (1979) 100.

\item M.~Campostrini, G.~Curci and A.~Pelissetto, \PLB{193} (1987) 279.
A.~Pelissetto, \APH{182} (1988) 177.

\item J.~Smit, \APP{17} (1986) 531. P.~Swift, \PLB{145} (1984) 256.
E.~Eichten and J.~Preskill, \NPB{268} (1986) 179.
 I.~Montvay, \PLB{199} (1987) 89; \NPBP{4}  (1988) 443.

\item I.~Montvay, in {\it Lattice 91}, \NPBP{26} (1992) 57.
J.~Smit, \NPBP{17} (1990) 3.

\item M.F.L.~Golterman, D.N.~Petcher and  J.~Smit, \NPB{370} (1992) 51.
M.F.L.~Golterman and D.N.~Petcher, \NPBP{26} (1992) 483.
W.~Bock, A.K.~De, C.~Frick, K.~Jansen and T.~Trappenberg,
\NPB{371} (1992) 683.
M.F.L.~Golterman and D.N.~Petcher \NPBP{26} (1992) 486.
M.F.L.~Golterman, D.N.~Petcher and E.~Rivas, Wash. U. preprint HEP/92-80,
\NPB{ } to appear.

\item D.B.~Kaplan, \PLB{288} (1992) 342; \NPBP{30} (1993) 597.

\item Y.~Shamir, Weizmann preprint WIS--93/56--JUNE--PH,
hep-lat/9306023, submitted to \PRL{ }

\item T.~Banks and A.~Dabholkar, \PRD{46} (1992) 4016.

\item N.~N.~Bogoliubov and D.~V.~Shirkov, {\it Introduction to the Theory
of Quantized Fields}, Interscience Publ. New York, 1959, p. 654.
H.~J.~Bremermann, R.~Oehme and J.~G.~Taylor, \PR{109} (1958) 2178.

\item S.G.~Krantz, {\it Function theory of Several Complex Variables},
John Wiley, New York, 1982, p. 133.

\item Y.~Shamir, in preparation.

\item J.M.~Rabin, \PRD{24} (1981) 3218.

\item J.~L.~Alonso, Ph.~Boucaud, J.~L.~Cortes and E.~Rivas, \MPL{5}
(1990) 275, \NPBP{17} (1990) 461. I-H.~Lee, \NPBP{17} (1990) 457.
S.~A.~Frolov and A.~A.~Slavnov, preprint MPI-Ph 93-12.
W.~Bock, J.~Smit and J.~C.~Vink, preprint ITFA 93-13, hep-lat 9306012.

\item A.~Borelli, L.~Maiani, G.-C.~Rossi, R.~Sisto and M.~Testa,
\NPB{333} (1990) 335.

\item Y.~Shamir, Weizmann preprint WIS-93/20/FEB-PH, to appear in \NPB.

\item L.S.~Adler W.~Bardeen, \PR{182} (1969) 1517

\item A.~Zee, \PRL{29} (1972) 1198

\end{list}

\end{document}